\documentclass[12pt,preprint,apj]{emulateapj}

\usepackage{epstopdf}
\usepackage{color}

\shorttitle{ Distances to NGC 4038/39 and NGC 5584}
\shortauthors{Jang \& Lee}

\begin{document}

\title{
THE TIP OF THE RED GIANT BRANCH DISTANCES TO TYPE IA SUPERNOVA HOST GALAXIES. III. NGC 4038/39 AND NGC 5584
}
 
\author{In Sung Jang and Myung Gyoon Lee }
\affil{Astronomy Program, Department of Physics and Astronomy, Seoul National University, Gwanak-gu, Seoul 151-742, Korea}
\email{isjang@astro.snu.ac.kr, mglee@astro.snu.ac.kr }


\begin{abstract}
We present the tip of the red giant branch (TRGB) distances to Type Ia supernova (SNe Ia) host galaxies NGC 4038/39 
and NGC 5584.
Based on the deep images constructed using archival Hubble Space Telescope data, 
we detect red giant branch stars in each galaxy.
$VI$ photometry of the resolved stars and corresponding $I$-band luminosity functions show the TRGB to be at $I_{TRGB}=27.67\pm0.05$ for NGC 4038/39 and $I_{TRGB}=27.77\pm0.04$ for NGC 5584.
From these estimates, we determine the distance modulus to NGC 4038/39 to be 
$(m-M)_0=31.67\pm0.05$ (random)$\pm0.12$ (systematic) 
(corresponding to a linear distance of $21.58\pm0.50\pm1.19$ Mpc)
and the distance modulus to NGC 5584 to be 
$(m-M)_0=31.76\pm0.04 $(random)$\pm0.12$ (systematic) 
(corresponding to a linear distance of $22.49\pm0.41\pm1.24$ Mpc).
We derive a mean absolute maximum magnitude of SNe Ia of $M_V=-19.27\pm0.08$ from the distance estimates of five SNe Ia (including two SNe in this study and three SNe Ia from our previous studies), and we derive a value of $M_V=-19.19\pm0.10$ using three low-reddened SNe Ia among the five SNe Ia.
With these estimates, we derive a value of the Hubble constant, $H_0 = 69.8\pm2.6 $(random)$ \pm3.9$ (systematic) km s$^{-1}$ Mpc$^{-1}$ and 
$72.2\pm3.3 $(random)$ \pm4.0$ (systematic) km s$^{-1}$ Mpc$^{-1}$, respectively. 
The value from the five SNe is similar to those from the cosmic microwave background analysis, 
and not much different within errors, from those of recent Cepheid calibrations of SNe Ia. 
The value from the three SNe is between the values from the two methods.
\end{abstract}


\clearpage

\section{Introduction}

\begin{figure*}
\centering
\includegraphics[scale=0.80]{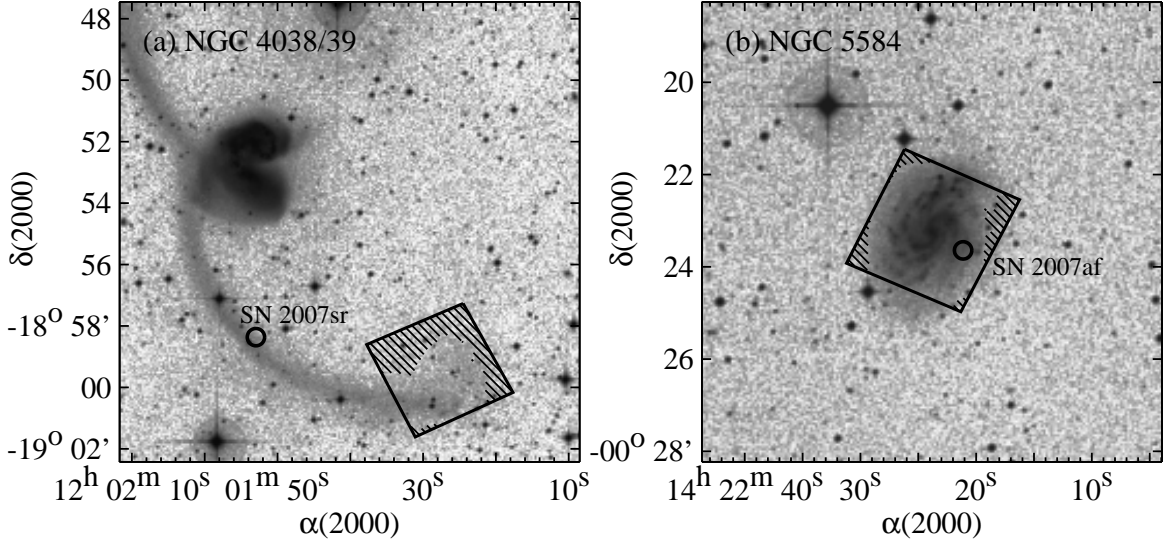} 
\caption{Identification of $HST$ fields for NGC 4038/39 (a) and NGC 5584 (b) overlaid on grey scale maps of the Digitized Sky Survey images. Only resolved stars in the hatched region of each field were used in the analysis. Locations of known SNe Ia are marked by open circles.
}
\label{fig_finder}
\end{figure*}

Type Ia supernovae (SNe Ia) are one of the most powerful distance indicators for distant galaxies in the cosmic expansion dominated field ($d\gtrsim100$ Mpc).
Three advantages of SNe Ia - a bright peak luminosity, a small luminosity dispersion, and a homogeneity in the universe - have enabled us to investigate the cosmic expansion 
rate ($H_0$) \citep{fre01, fre12,san06,rie11} and the cosmic acceleration rate ($\Lambda$) \citep{rie98,per99}. Although great progress has been made over the last two decades, more accurate calibration of peak luminosity of SNe Ia is needed to derive more accurate values of cosmological parameters.

Recently, two large projects, Supernovae and $H_0$ for the Equation of State (SH0ES, \citep{rie11}) and Carnegie Hubble Program (CHP, \citep{fre12}), have been working on the calibration of SNe Ia to improve the measurement of  cosmological parameters. 
They performed the luminosity calibration of SNe Ia based on the Cepheid distance estimates of 8 SNe Ia host galaxies and derived similar values of the Hubble constant: $74.8\pm3.1$ km s$^{-1}$ Mpc$^{-1}$\citep{rie11} and $74.3\pm2.1$ km s$^{-1}$ Mpc$^{-1}$ \citep{fre12}.

However, these values of the Hubble constant are somewhat larger than those
based on recent analysis of the cosmic microwave background radiation (CMBR).  
\citet{ben13}  derived the cosmological parameters from the analysis of the angular power spectrum of CMBR in WMAP9 assuming a flat $\Lambda$CDM cosmology, and presented $H_0 = 69.32 \pm 0.80$ km s$^{-1}$ Mpc$^{-1}$.
Later, the Planck Collaboration lead to an even smaller value of the Hubble constant, $H_0 = 67.8 \pm 0.9$ km s$^{-1}$ Mpc$^{-1}$, using new CMBR data from the Planck satellite \citep{pla14,pla15}.
Thus the traditional distance ladder method and the CMBR analysis based on $\Lambda$CDM cosmology show a 2--3$\sigma$ level of difference in the estimates of the Hubble constant \citep{ben14}. 
This is called "the Hubble tension", one of the most critical issues in modern cosmology.
In addition, recent studies of the baryon acoustic oscillation (BAO) in the range of redshift combined with SNe Ia data presented similar values to those from the CMBR analysis \citep{aub14}.

We have been working on the luminosity calibration of SN Ia  by measuring the precise distances to SN Ia host galaxies based on the Tip of the Red Giant Branch (TRGB) \citep{lee93}.
The TRGB method uses old red giant branch (RGB) stars, classified as population II. 
The metallicity dependence of $I$-band luminosities of TRGB stars is known to be small \citep{lee93, bel01, riz07, mad09, sal97}.
Moreover, the TRGB method often uses RGB stars located in the halo, where the internal extinction is expected to be negligible and the crowing of stars is much less than the disk or the bulge. 
\citet{lee12} 
(Paper I)
 measured the TRGB distance to M101 hosting SN 2011fe, which is the nearest SN Ia with modern CCD photometry. 
\citet{lee13} 
(Paper II) 
determined the TRGB distances to M66 and M96 in the Leo I Group, hosting SN 1989B and SN 1998bu. By combining the TRGB distance estimates of three SNe Ia from Papers I and II, and optical light curves for three SNe Ia, we derived the Hubble constant, $H_0 = 68.4 \pm 2.6 $(random)$ \pm3.7 $(systematic) km s$^{-1}$ Mpc$^{-1}$ \citep{lee13}, which is slightly smaller than values from the Cepheid calibrated SNe Ia \citep{rie11, fre12}. 

In this paper, we present the TRGB distances to two additional SN Ia host galaxies, NGC 4038/39 hosting SN 2007sr and NGC 5584 hosting SN 2007af.
NGC 4038/39 are a well known pair of galaxies showing early stage merger of disk galaxies.
In 2007, a supernova was discovered near the southern tidal tail of this galaxy, SN 2007sr \citep{dra07}.
Spectroscopic observations confirmed that it is a SN Ia \citep{nai07,umb07}. 
Photometric follow-up observations were made from optical to near-infrared \citep{sch08, hic09}.
Light curve fits on this SN suggested that its internal reddening is expected to be small, $A_V \sim 0.2$ \citep{sch08,phi13}.

There are three distance estimates to NGC 4038/39 based on the TRGB in the literature. \citet{sav04} analysed Hubble Space Telescope $(HST)$/Wide Field Planetary Camera 2 $(WFPC2)$ images of a field on the southern tidal tail and derived a TRGB distance, $(m-M)_0 = 30.70\pm0.25$ ($d = 13.8 \pm1.7$ Mpc). This value is significantly smaller than previous estimates based on radial velocities of NGC 4038/39 ($d\sim20$ Mpc ($V_{Local\ Group} \sim 1400$ $km$ $s^{-1}$, assuming $H_0 \sim 70$ km s$^{-1}$ Mpc$^{-1}$)). Later, \citet{sav04} presented a similar distance estimate, $(m-M)_0 = 30.62\pm0.17$ ($d = 13.3 \pm1.0$ Mpc), using the new $HST$/Advanced Camera for Surveys (ACS) image data \citep{sav08}. 
However, \citet{sch08} presented a significantly larger value, $(m-M)_0 = 31.51\pm0.17$ ($d = 20.0 \pm1.6$ Mpc), based on the same ACS image data used in \citet{sav08}. 
On the other hand, \citet{rie11} detected Cepheid variables in the main bodies of NGC 4038/39 and derived a distance modulus, $(m-M)_0 = 31.66\pm0.08$ ($d = 21.5 \pm0.6$ Mpc). This is similar to the value in \citet{sch08}, but much larger than the value given by \citet{sav08}.

NGC 5584 is a moderately inclined ($\theta=42.4\arcdeg$) spiral galaxy hosting SN 2007af, which is one of the most well monitored SNe Ia. Before its pre-maximum, spectroscopic and photometric observations from optical to near infrared have been made \citep{ham06,hic09}. Moreover the internal reddening for SN 2007af is known to be small, $A_V=0.39\pm0.06$ \citep{sim07}. Thus, SN 2007af is an ideal target for the luminosity calibration of SNe Ia.
Previous distance estimates based on the Tully-Fisher relation show a wide range of values with large uncertainty: $(m-M)_0=31.12 \sim 31.59$ with uncertainties of $\sim0.4$ \citep{the07,spr09,sor12}.
\citet{rie11} detected 94 Cepheid variables based on the optical/near-infrared image data taken with $HST$/WFC3, and derived a distance, $(m-M)_0=31.72 \pm 0.07$ ($d = 22.1 \pm0.7$ Mpc). This value is much larger than those based on the Tully-Fisher relations. To date, there is no TRGB distance estimate for NGC 5584.

This paper is composed as follows. Section 2 describes data  reduction. Section 3 presents color-magnitude diagrams and distance estimates for each galaxy. Implications of our results are discussed in Section 4. 

\section{Data and Data Reduction}

\begin{deluxetable}{lcc}
\setlength{\tabcolsep}{0.05in}
\tablecaption{A summary of $HST$ observations  of NGC 4038/39 and NGC 5584}
\tablewidth{0pt}

\tablehead{ \colhead{ } & \colhead{NGC 4038/39} &  \colhead{NGC 5584}
}
\startdata
R.A.(2000) 	& $12^h01^m27.^s45$ 	& $14^h22^m23.^s62$   \\
Dec(2000) 	& $-18\arcdeg59\arcmin28\farcs7$  	& $-00\arcdeg23\arcmin12\farcs7$  \\
Prop ID. 	& 10580		 	& 11570  \\
Detector	& ACS/WFC		& WFC3/UVIS \\
Exposure, V & 10,870s, $F606W$ & 45,540s, $F555W$ \\
Exposure, I	& 8,136s, $F814W$ & 14,400s, $F814W$ \\

\enddata
\label{observation}
\end{deluxetable}

\begin{figure*}
\centering
\includegraphics[scale=0.70]{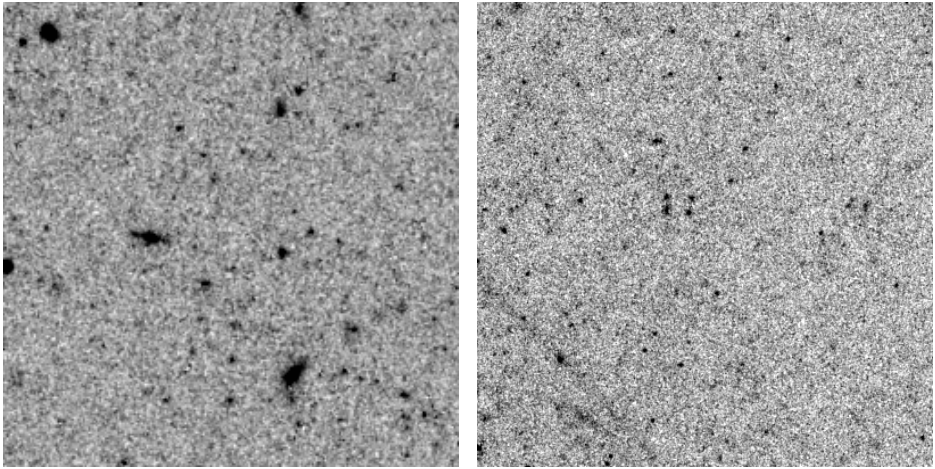} 
\caption{$5\arcsec \times 5\arcsec$ sections of reconstructed $F814W$ images for NGC 4038/39 (left) and NGC 5584 (right). The pixel scales of each image are $0\farcs04$ and $0\farcs025$, respectively. Both images clearly show point-like and extended sources.
}
\label{fig_resolve}
\end{figure*}

\begin{deluxetable*}{lcccc}
\setlength{\tabcolsep}{0.05in}
\tablecaption{A summary of $HST$ observations  of NGC 2419 and 47 Tuc from the archive}
\tablewidth{0pt}

\tablehead{ \colhead{ } & \multicolumn{2}{c}{NGC 2419} &  \multicolumn{2}{c}{47 Tuc} \\
	& ACS/WFC	& WFC3/UVIS &	ACS/WFC&	WFC3/UVIS }
\startdata
R.A.(2000) 	& $07^h38^m08.^s27$ 	& $07^h38^m08.^s54$ & $00^h22^m38.^s52$ & $00^h22^m35.^s92$ \\
Dec(2000) 	& $38\arcdeg51\arcmin36\farcs3$  	& $38\arcdeg52\arcmin40\farcs6$  & $-72\arcdeg04\arcmin01\farcs1$ & $-72\arcdeg04\arcmin05\farcs3$ \\
Prop ID. 	& 9666		 	& 11903 & 10048, 11677 & 11903, 11452\\
Exposure, $F555W$ & 720s  & 1,160s & 2,139s & 1,160s\\
Exposure, F606W & 676s  &  800s & 163,676s &  18,376s\\
Exposure, $F814W$	& 676s  & 1,300s & 184,242s & 3,400s\\
\enddata
\label{observation_gc}
\end{deluxetable*}

The image data for NGC 4038/39 and NGC 5584 were acquired from the $HST$ archive. 
Table \ref{observation} lists the information for the $HST$ image data used in this study.
The image data for NGC 4038/39 covering the southern tidal tail, as shown in Figure \ref{fig_finder} (a), were taken with the ACS in Wide Field Channel (WFC) mode (Proposal ID: 10580) with exposure times of $4$ $\times \sim2700$s for $F606W$ and $3$ $\times \sim2700$s for $F814W$. 
Corresponding total exposure times were 10870s and  8136s for $F606W$ and $F814W$, respectively.
Charge transfer efficiency (CTE) corrected and flat fielded images (indicated by suffix \_flc.fits) were combined to make a single drizzled image using the DrizzlePac\footnote{http://drizzlepac.stsci.edu/}.
 Because individual \_flc images were severely contaminated by cosmic rays, the automatic image aligning tool, TweakReg task, failed to provide fine solutions.
 Thus, we used an alternative strategy, as described in the following.
 
 First, we performed an aperture photometry on the \_flc images with aperture radii of  $0\farcs040$ and $0\farcs125$ using DAOPHOT in the IRAF package. 
 The magnitude difference between these two aperture radii is defined as the concentration index, $C$.
 Next, sources with $0.9 < C \leq 1.6$ were selected.
 Through this process, sources with abnormally narrow (e.g., cosmic rays) or abnormally broad (e.g., background galaxies) radial profiles were rejected.
 Second, we converted image coordinates for the selected sources to the World Coordinate System (WCS) based on the information listed in the header.
 The WCS coordinates for a pair of \_flc images were matched with a matching criteria of 0\farcs05 (1 pixel).
 This process rejects most non-stellar objects such as cosmic rays and hot pixels.
Finally, the files to be used with the TweakReg task were created. 
These files contain source coordinates and flux counts.
Using these files, the TweakReg task output better solutions of $\Delta X$ and $\Delta Y$ with a rms value of $\sim0.06$ pixel ($0\farcs003$).\\
The aligned images were combined into a single drizzled images using the AstroDrizzle task.
The final\_pixfrac and final\_scale values used are 
1.0 and $0\farcs04/$pixel, respectively.
Output images show FWHM of 2.64 pixels ($0\farcs106$) for $F606W$ and 2.38  pixels ($0\farcs095$) for $F814W$.

Observations for NGC 5584 were done with the Wide Field Camera 3 (WFC3) in Ultraviolet-Visible (UVIS)  mode with the primary aim of detecting Cepheid variables (Proposal ID : 11570).
Centered on NGC 5584, the $2\farcm7 \times 2\farcm7$ field covering the main body and inner halo of the galaxy, as shown in Figure \ref{fig_finder} (b), was observed 
with exposure times of $76$ $\times \sim600$s for $F555W$ and $24$ $\times \sim600$s for $F814W$. 
Total exposure times are 45,540s and  14,400s for $F555W$ and $F814W$, respectively.
The \_flc images were made from the flat-fielded images (indicated by suffix \_flt.fits) and the raw images (indicated by suffix \_raw.fits) 
by applying pixel-based CTE correction software\footnote{http://www.stsci.edu/hst/wfc3/ins\_performance/CTE/}.
Derived \_flc images were aligned in the same way as was done for NGC 4038/39 images. 
Mean rms values for $\Delta X$ and $\Delta Y$ from the TweakReg task are 
$\sim0.08$ pixels ($\sim0\farcs0032$) for both filters.
Aligned \_flc images were then combined using the AstroDrizzle task with a final\_pixfrac value of 0.7 and a final\_scale value of $0\farcs025/$pixel. 
RMS values of weight images for each filter are 
smaller than 4\% of the median value.
FWHMs for point sources in the final drizzled images are 3.00 pixels ($0\farcs075$) for $F555W$ and 3.22 pixels ($0\farcs081$) for $F814W$.
Figure \ref{fig_resolve} shows $5\arcsec\times5\arcsec$ sections of the output $F814W$ images of NGC 4038/39 (a) and of NGC 5584 (b). 
Point sources and extended sources (mostly background galaxy) can be clearly seen in the images. 

\begin{figure*}
\centering
\includegraphics[scale=0.80]{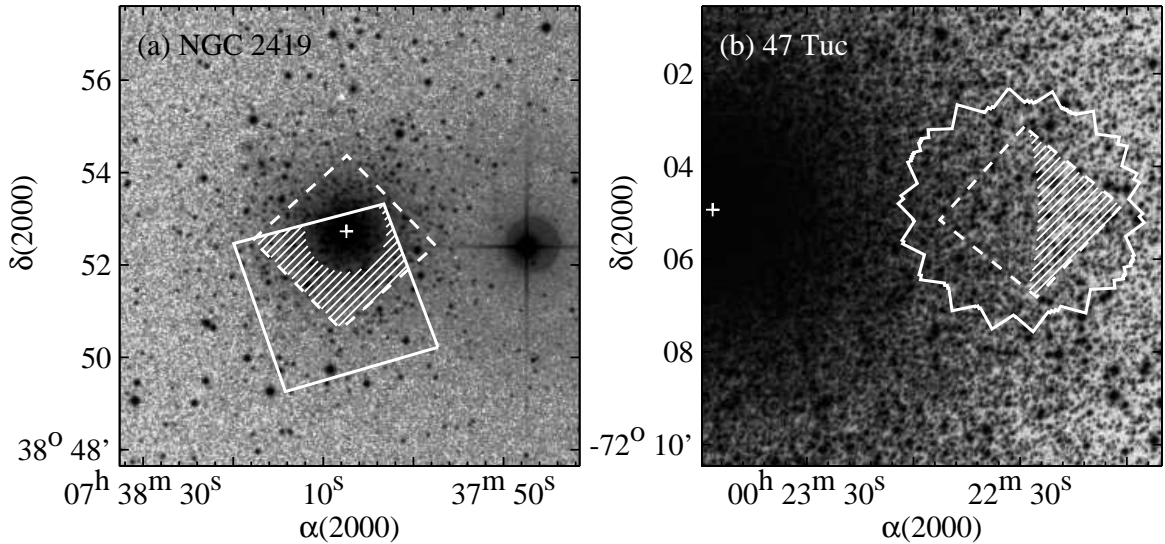}
\caption{Finding charts for ACS/WFC fields (solid square in (a) and polygon in (b)) and WFC3/UVIS  fields (dashed-line squares) of  NGC 2419 (a) and 47 Tuc (b) overlaid on Digitized Sky Survey images.
The center of each star cluster is indicated by a white cross. 
Only point sources in the hatched region of each cluster ($R>1\arcmin$ for NGC 2419 and $R>7\arcmin$ for 47 Tuc) were used for the photometric transformation.
}
\label{fig_finder_gc}
\end{figure*}

 PSF photometry was done on the drizzled images using the standard routine : Daofind, Phot, and Allstar tasks in IRAF/DAOPHOT \citep{ste87}. Initial aperture photometry with aperture radii of $0\farcs040$ and $0\farcs125$ were done to derive the concentration index for each stellar object. 
 With this combination of aperture radii, point sources show mean concentration index values of $C=1.25$ for NGC 4038/39 and $C=1.05$ for NGC 5584.
PSF images were constructed for each galaxy based on $\sim50$ bright, isolated, and $C \sim1.25$ or $1.05$ stellar objects.
Source coordinates to be used in the photometry were determined from $F814W$-band images for each galaxy with a detection threshold of 2.
We used revised photometric zeropoints for ACS/WFC 
(26.416 for $F606W$ and 25.529 for $F814W$) 
and WFC3/UVIS
(25.81 for $F555W$ and 24.67 for $F814W$), provided by STScI
\footnote{http://www.stsci.edu/hst/acs/analysis/zeropoints/zpt.py}\footnote{http://www.stsci.edu/hst/wfc3/phot\_zp\_lbn}. 
Aperture correction values to the aperture radii of $0\farcs5$ (for ACS/WFC) or $0\farcs4$ (for WFC3/UVIS) were calculated by checking the curve of growth for bright isolated stars.
Derived values are  
$-0.162$, $-0.157$ for $F606W$, $F814W$ on ACS/WFC and 
$-0.139$, $-0.229$ for $F555W$, $F814W$ on WFC3/UVIS with an uncertainty of 0.02.
Additional aperture corrections up to the infinite aperture radius were also applied.
Corresponding values derived from the enclosed energy curves provided by STScI\footnote{http://www.stsci.edu/hst/acs/analysis/apcorr} \footnote{http://www.stsci.edu/hst/wfc3/uvis\_ee\_model\_smov.dat} are :  
$-0.095, -0.098$ for $F606W$, $F814W$ on ACS/WFC and 
$-0.102, -0.107$ for $F555W$, $F814W$ on WFC3/UVIS, respectively.
Finally, ACS magnitudes were converted to standard Johnson-Cousins magnitudes using the photometric transformation equations listed in \citet{sir05}.

\subsection{Photometric Transformations}

 \begin{figure*}
\centering
\includegraphics[scale=0.80]{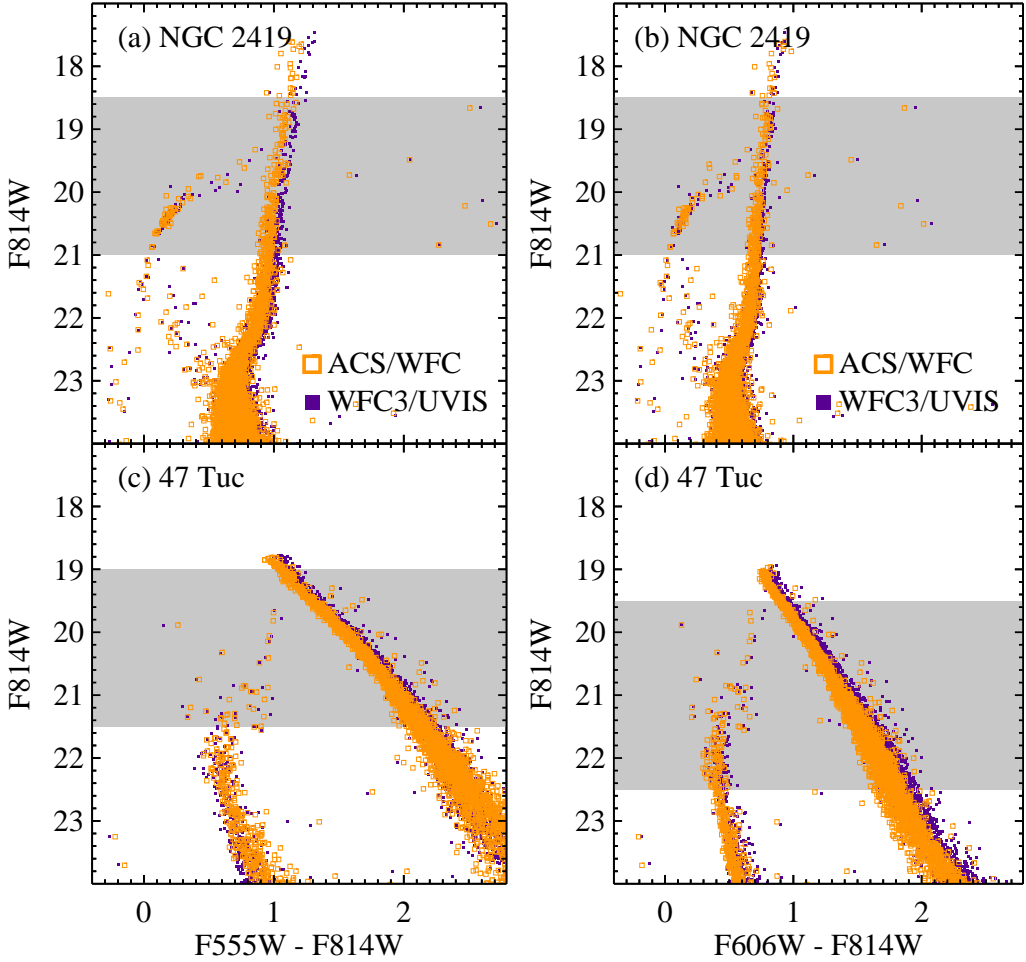}
\caption{ 
Color magnitude diagrams of resolved stars in NGC 2419 (a and b) and 47 Tuc (c and d). Open and filled squares indicate the photometric data from the ACS/WFC and the WFC3/UVIS, respectively. We used bright and non-saturated stars, which are located in the shaded boxes in each panel, for the photometric transformation.
}
\label{fig_cmd_gc}
\end{figure*}  

\begin{figure*}
\centering
\includegraphics[scale=0.80]{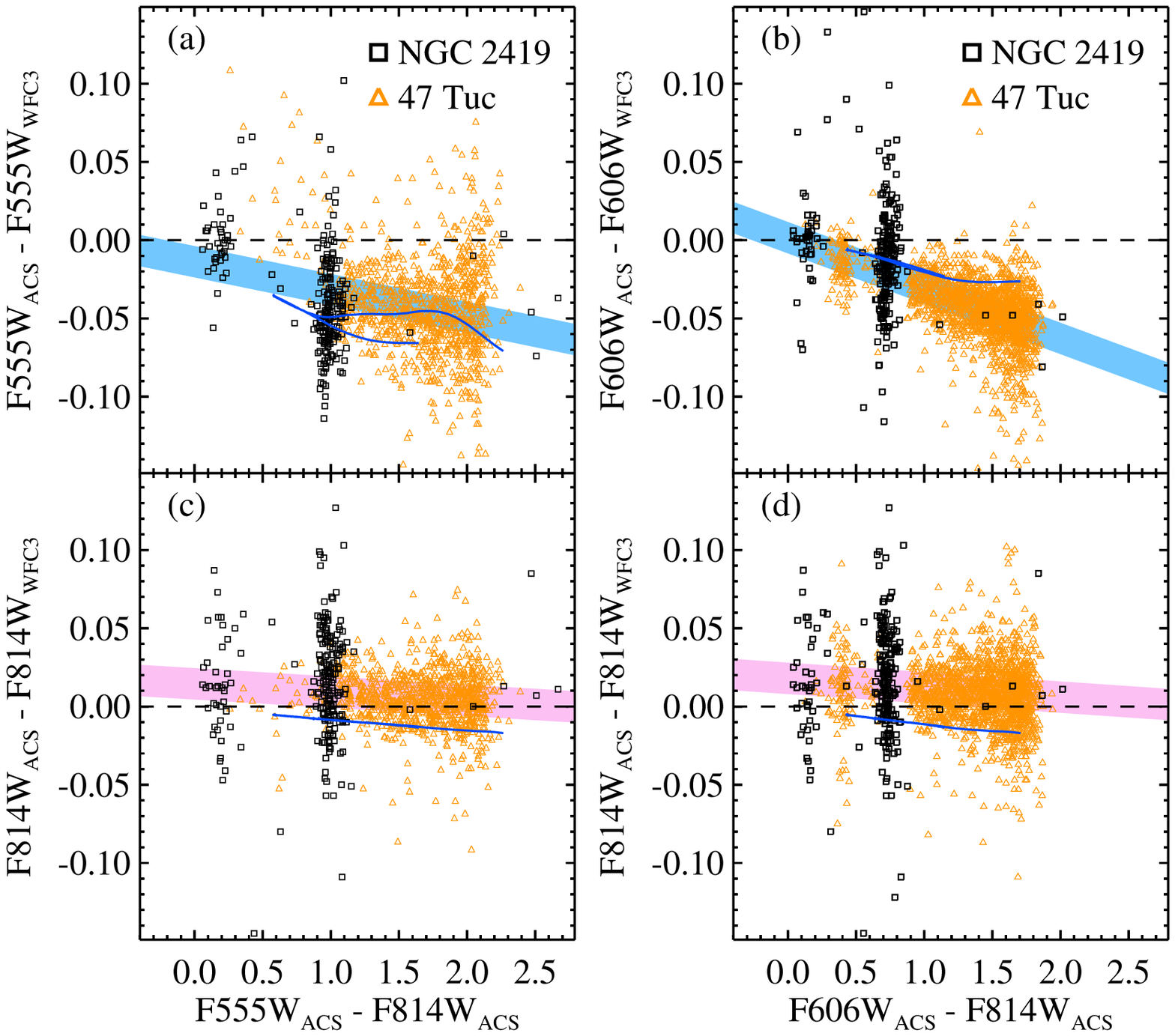} 
\caption{
((a) and (c)) Comparison of photometric systems between ACS/WFC and WFC3/UVIS for $F555W$-band, and $F814W$-band in terms of $F555W-F814W$ color in the ACS/WFC system. ((b) and (d)) Same as (a) and (c) but for the (F606W)-band.
Open squares and open triangles denote detected stars in NGC 2419 and 47 Tuc, respectively.  Solid lines in each panel indicate the  stellar isochrones for age=12 Gyr and  [Fe/H]$=-1.4$, which is the mean metallicity of two globular clusters, provided by the Dartmouth group \citep{dot08}. Shaded regions indicate the $1 \sigma$ uncertainty area of the photometric transformation.
}
\label{fig_transformation}
\end{figure*} 

The photometric systems for ACS/WFC and WFC3/UVIS are not the same.
There is no TRGB calibration for the WFC3/UVIS system yet.
To apply the TRGB calibration, and to facilitate comparisons of the two galaxies in this study and those in our previous studies (\citet{lee12, lee13}), we derived photometric transformations from the WFC3/UVIS system to the ACS/WFC system as follows.
 We used archival image data for two Milky Way globular clusters, NGC 2419 and 47 Tuc, which were observed using both ACS/WFC and WFC3/UVIS, as listed in Table \ref{observation_gc} and shown in Figure \ref{fig_finder_gc}.
 These two globular clusters have been used for the photometric transformation from the ACS/WFC system to the Johnson-Cousin system in \citet{sir05}, \citet{har07}, and \citet{sah11}.
 We calibrated raw image data to construct CTE corrected drizzled images and carried out PSF photometry as described in Section 2. We then matched the photometric catalogues from ACS/WFC and WFC3/UVIS image data.

 Constructed color-magnitude diagrams (CMDs) for resolved stars in NGC 2419 and 47 Tuc are shown in Figure \ref{fig_cmd_gc}.
To reduce the uncertainty due to the stars in the crowded field, we used only the stars located in relatively low density 
environments at $R_{NGC2419} \geq 1\arcmin$ and $R_{47 Tuc} \geq 7\arcmin$, indicated by the hatched lines in Figure \ref{fig_finder_gc}.
CMDs from ACS/WFC data (open squares) and WFC3/UVIS data (filled squares) 
look very similar.
Stars with $F555W-F814W \gtrsim0.8$, WFC3/UVIS data show in general redder colors than ACS/WFC data. 
For blue stars with $F555W-F814W \sim 0.0$, however, WFC3/UVIS data shows slightly bluer color than ACS/WFC data. This trend is 
seen in both globular clusters.
In order to compare the difference 
quantitatively, we selected bright and non-saturated stars in the shaded regions ($18.5 < F814W_{ACS} < 21.0$ for NGC 2419, and $19.0 < F814W_{ACS} < 21.5$ or $19.5 < F814W_{ACS} < 22.5$ for 47 Tuc) of Figure \ref{fig_cmd_gc} and compared them in Figure \ref{fig_transformation}.

 Figure \ref{fig_transformation} shows comparisons of magnitudes between the two photometric systems in terms of their colors. $F555W_{ACS} - F814W_{ACS}$ colors range from 0.0 to 2.2, and their mean offsets are $\sim-0.04$ for $F555W$ and $\sim0.01$ for $F814W$. 
 We plot isochrones, as a reference, for the 12 Gyr age and $[Fe/H]=-1.4$, which is a mean metallicity of NGC 2419 and 47 Tuc,
 provided by Dartmouth group \citep{dot08}. Data points and stellar isochrones are in good agreement, mean offsets are estimated to be $\sim$ 0.01  in both filters. 
 We derive a photometric transformation as follows : \\

$F555W_{ACS}$ = $F555W_{WFC3}-(0.0137\pm0.0044) - (0.0178\pm0.0030)(Color)$, \\
and \\
$F814W_{ACS}$ = $F814W_{WFC3} + (0.0145\pm0.0036) - (0.0053\pm0.0024)(Color)$ \\
\noindent where $Color = F555W_{ACS}-F814W_{ACS}$.\\
Note that the amount of correction is very small: 0.05 mag in $F555W$ and smaller than 0.01 mag in $F814W$ for $Color \sim 2.0$. 

We applied this photometric transformation to the photometry of NGC 5584.
Transformation uncertainties are estimated to be $\sim0.01$ mag considering uncertainties of aperture corrections, although the fitting uncertainties are much smaller than 0.01 mag.
 
We derived the same photometric transformation for $F606W$-band for the future studies.\\
$F606W_{ACS}$ = $F606W_{WFC3}+(0.0016\pm0.0021) - (0.0322\pm0.0019)(Color)$, \\
and \\
$F814W_{ACS}$ = $F814W_{WFC3} + (0.0156\pm0.0023) - (0.0060\pm0.0020)(Color)$ \\
\noindent where $Color = F606W_{ACS}-F814W_{ACS}$.

\subsection{Artificial star test} 
 
\begin{figure*}
\centering
\includegraphics[scale=0.90]{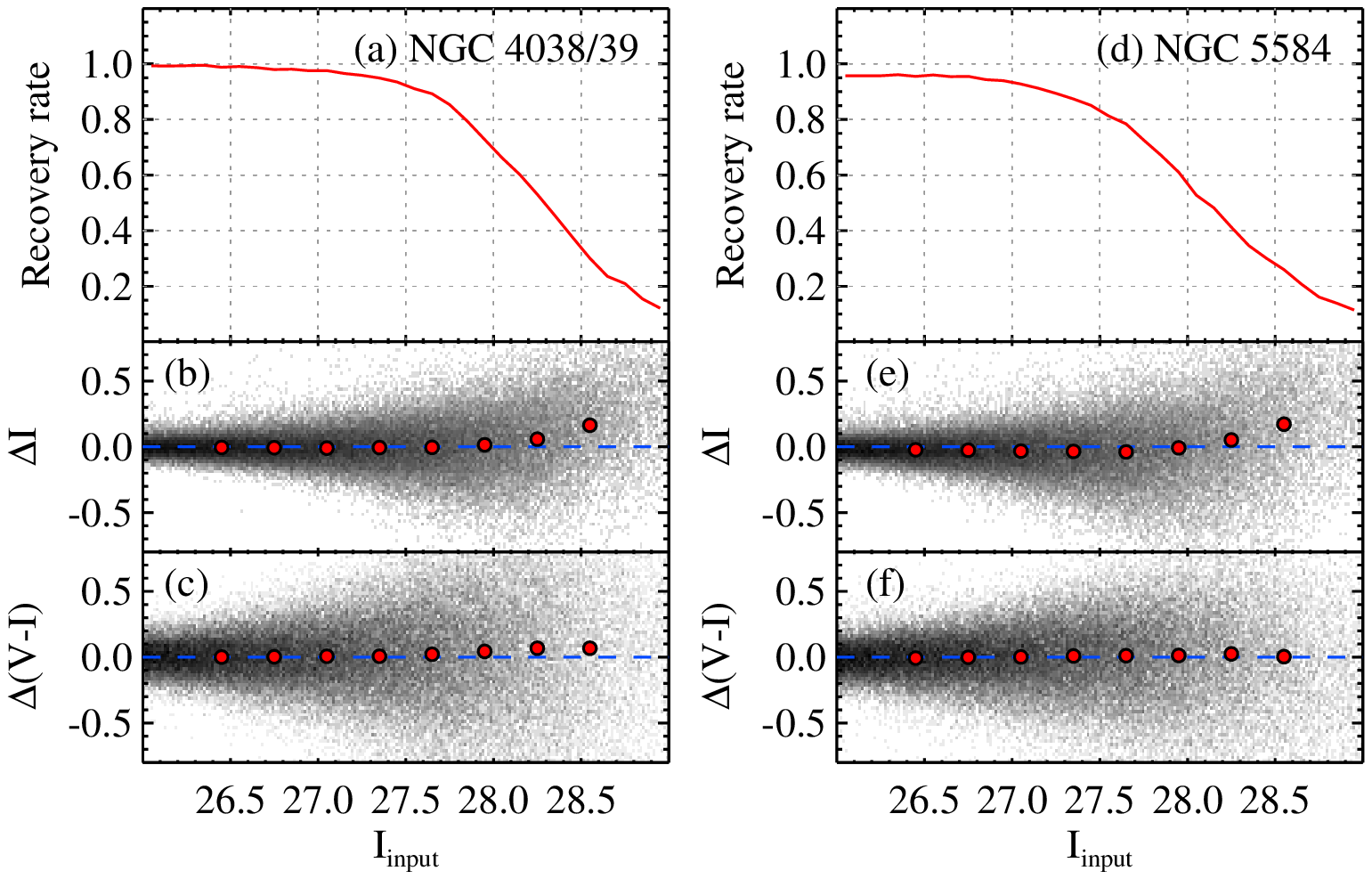} 
\caption{ (a) Recovery rate of the PSF photometry as a function of $I$-band magnitude for NGC 4038/39. (b,c) Random and systematic uncertainties of $I$-band magnitudes (b) and $(V-I)$ colors (c), estimated from artificial star experiments. The differences between input and output magnitudes ($\Delta I = I_{input} - I_{output}$) and colors ($\Delta (V-I) = (V-I)_{input} - (V-I)_{output}$) are shown as density maps, and their median offsets in each magnitude bin are shown as larger circles. ((d), (e) and (f)) Same as (a), (b), and (c) but for NGC 5584. 
}
\label{fig_comp}
\end{figure*}  
 
We checked the photometric completeness and systematic errors of our photometry through the artificial star experiment. 
To do this, we selected test fields with a size of $0\farcm5\times1\farcm0$ from the hatched regions in Figure \ref{fig_finder}.
Next, we added about $150$ and $300$ artificial stars into the test fields of NGC 4038/39 and NGC 5584, which correspond to $\sim10$\%  of the total amount of stars detected in each field.
These added stars were uniformly distributed throughout the fields and have a magnitude range of $26 \leq I \leq 29$.
We set $(V-I) \sim 1.7$ for input artificial stars in order to make them similar to the estimated median $(V-I)$ color at the TRGB magnitude (see Section 3.2).
Then, we preformed PSF photometry on the fields in the same way as done on the original images. 
We then checked recovery rates and photometric systematics.
We iterated this process 600 and 300 times for NGC 4038/39 and NGC 5584, respectively, to keep a larger number of artificial stars and to reduce statistical uncertainties. In total, about $100,000$ artificial stars were used for each field in the test.

Figures \ref{fig_comp} (a) and (d) show the recovery rates for artificial stars as a function of input $I$-band magnitudes for these two galaxies. 
About 90\% and 70\% of stars are expected to be recovered at the TRGB magnitude of NGC 4038/39 ($I_{TRGB} = 27.67\pm0.05$) and NGC 5584 ($I_{TRGB} = 27.77\pm0.04$).
The 50\% recovery rates are estimated to be $I\sim28.3$ for NGC 4038/39 and $I\sim28.1$ for NGC 5584 . 
It is noted that, although exposure times of NGC 4038/39 are much shorter than those of NGC 5584, the recovery rates are higher in the same magnitude range.
This can be explained as follows. 
First, the field used in the analysis of NGC 4038/39 is much farther from the center of the host galaxy than that of NGC 5584, so that the expected stellar density is much lower.  
Second, NGC 4038/39 was 
observed with ACS/WFC, which has 
higher sensitivity than the WFC3/UVIS at longer wavelengths of $\sim$ 400 nm. 
Third,  NGC 4038/39 was 
observed with $F606W$-band, which has a much wider spectral window than the $F555W$-band used for NGC 5584.

Figures \ref{fig_comp} (b), (c), (e) and (f) show random uncertainties and systematic offsets of $I$-band magnitudes ($\Delta I = I_{input} - I_{output}$) and $V-I$ colors ($\Delta (V-I) = (V-I)_{input} - (V-I)_{output}$) for the two galaxies. 
Random uncertainties increase gradually, 
being proportional to the input $I$-band magnitudes.
Systematic offsets are noted. 
For $I_{input} \gtrsim 28.0$ in (b) and (e), $\Delta I$ values are positive. 
This means that those faint stars are measured slightly brighter than their intrinsic brightness.
At the TRGB magnitude of NGC 4038/39 and NGC 5584, such magnitude and color offsets are smaller than 0.02 mag, so we ignored these offsets.

\section{Results}

\subsection{Color-magnitude Diagrams} 

The $HST$ fields used in this study cover the tidal tail in NGC 4038/39, and the disk and bulge in NGC 5584. 
Therefore the color-magnitude diagrams of these fields 
show various populations of stars, such as young main sequence, blue and red super giant, asymptotic giant branch (AGB) and  RGB stars.
These mixed populations increase uncertainties in determining the TRGB.
In order to reduce the contamination due to multiple stellar populations and to sample the RGB population as much as possible, we selected regions with the lowest sky background level, avoiding spiral arms and tidal tails. Chosen regions are marked by the hatched regions in Figure \ref{fig_finder}.

Figure \ref{fig_cmd} show the CMDs of resolved stars in the selected regions of NGC 4038/39 (a), and NGC 5584 (b). 
We calculated the expected number of galactic foreground stars in each CMD using the TRILEGAL program  
provided by the Padova group \citep{gir12}. 
Derived numbers of stars are smaller than ten for both figures, indicating that a contamination by the foreground stars is negligible.
We also checked the background galaxy contamination by comparing our CMD with the CMD of the Hubble eXtreme Deep Field (HXDF), where background galaxies are dominated.
We used $F606W$ and $F814W$ combined images provided by \citet{ill13}.
Total exposure times are 174,400s for $F606W$, and 50,800s for $F814W$.
We then carried out the PSF photometry in the same method as for NGC 4038/39 and NGC 5584.
Constructed CMD for the point sources in the entire field of HXDF is shown in Figure \ref{fig_cmd} (c). Most of sources are expected to be unresolved background galaxies.
We then count the number of sources inside the slanted boxes of each panel, which will be used in the TRGB determination (see next section) : 5247 sources for NGC 4038/39, 4491 sources for NGC 5584, and 318 sources for HXDF. After the correction for the field area differences (4.9 arcmin$^2$ for NGC 4038/39 field, 1.0 arcmin$^2$ for NGC 5584 field, and 10.8 arcmin$^2$ for HXDF), we conclude that the background galaxy contaminations in the CMDs are very small : $\sim3$\% for NGC 4038/39 and $\sim1$\% for NGC 5584. We ignored this contamination in the following analysis.

The CMDs for NGC 4038/39 and NGC 5584 show clear RGB populations as well as relatively weak main sequence and AGB populations. 
Both CMDs show a clear number density enrichment of RGB stars at  $I\sim27.7$ for NGC 4038/39, and $I\sim27.8$ for NGC 5584, which correspond to the TRGB.

\subsection{Distance Estimation}

\begin{figure*}
\centering
\includegraphics[scale=0.80]{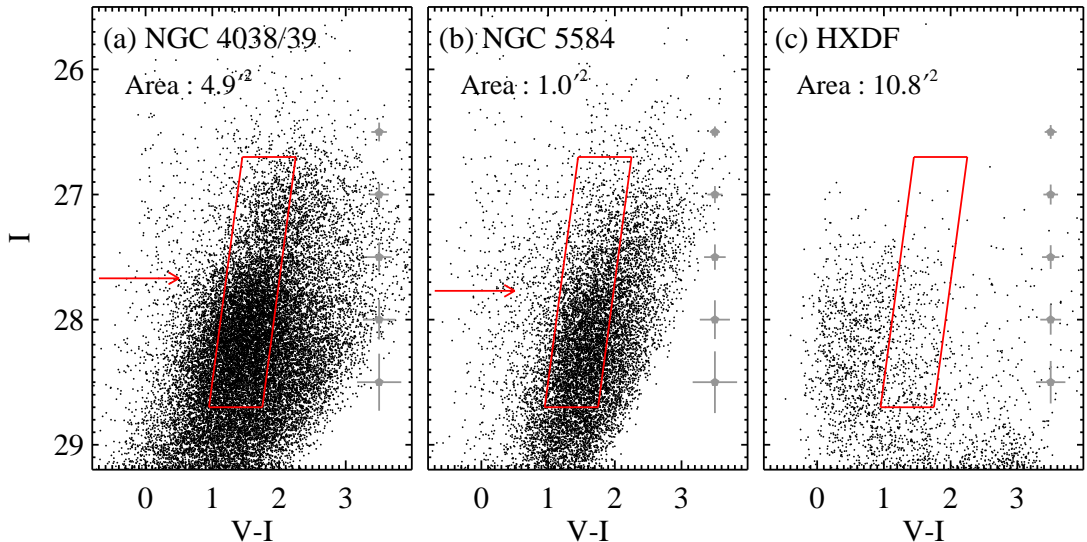} 
\caption{Color-magnitude diagrams of resolved stars in the selected regions of NGC 4038/39 (a) and NGC 5584 (b). We also showed the CMD of point sources in the Hubble eXtreme Deep Field (HXDF) as a reference (c). Slanted boxes indicate the boundary of resolved stars used in the TRGB analysis. Estimated TRGB magnitudes are marked by arrows. Crosses with error bars represent mean photometric errors.
}
\label{fig_cmd}
\end{figure*}

Our photometry of NGC 4038/39 and NGC 5584 are deep enough to reach more than one magnitude below the TRGB.
This enables us to measure distances with the TRGB, a known precise distance indicator \citep{lee93, fre10}.
We derived $I$-band luminosity functions for resolved stars inside the slanted boxes of Figure \ref{fig_cmd}, designed to preferentially select the blue RGB stars.\\
The use of the blue RGB stars in the analysis has several advantages, as noted in \citet{jan14}.
First, the TRGB for the bluer RGB stars is less curved than that of the redder RGB stars in the $I - (V-I)$ CMD.
Second, the bluer RGB stars have a relatively higher photometric recovery rate than the redder RGB stars.
Third, the bluer RGB stars are relatively less affected by the host galaxy reddening.

\begin{figure*}
\centering
\includegraphics[scale=0.75]{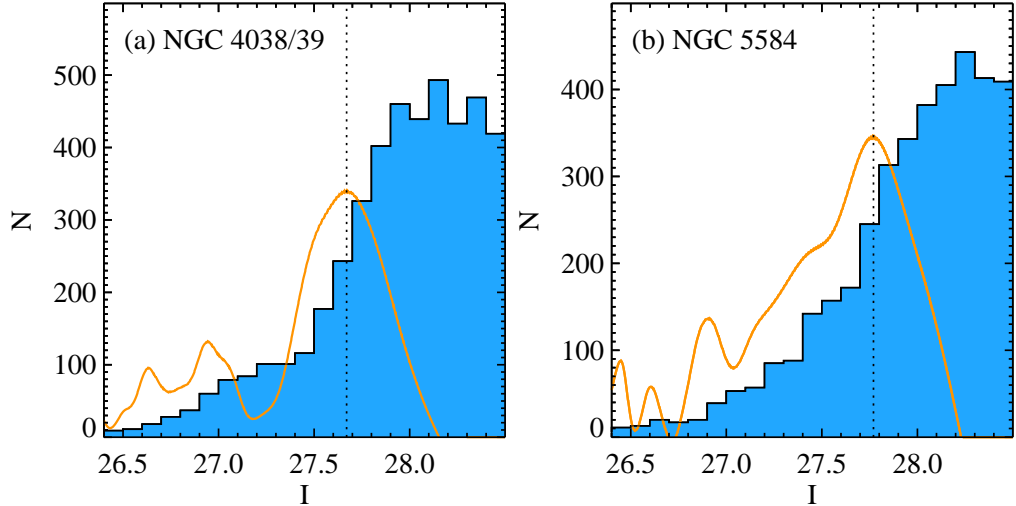} 
\caption{$I$-band luminosity functions for the resolved stars in the slanted boxes in Figure \ref{fig_cmd} of NGC 4038/39 (a) and NGC 5584 (b). 
Edge detection responses and the estimated TRGB magnitudes for each galaxy are shown by solid and dotted lines, respectively.
}
\label{fig_trgb}
\end{figure*}

Figure \ref{fig_trgb} shows derived $I$-band luminosity functions and corresponding edge-detection responses.
We used the logarithmic form of edge-detection algorithm \citep{sak96, men02}, which is described by 
\begin{displaymath}
E(I) = \sqrt\Phi[log[\Phi(I+\sigma_I)]-log[\Phi(I-\sigma_I)]]
\end{displaymath}
\noindent
, where $\Phi(I)$ is gaussian-smoothed luminosity function and $\sigma_I$ is the mean photometric error.
 
Edge-detection response functions show strong peaks at $I\sim27.7$ for NGC 4038/39 and $I\sim27.8$ for NGC 5584, as well as relatively weak peaks at between $I=26.4$ and $27.0$.
In a galaxy that contain various stellar populations, the number of AGB stars is much smaller than that of the RGB stars. 
Therefore, it is expected that the luminosity function of the red giant stars shows a weak jump at the AGB tip, and a much stronger jump at the RGB tip.
We determined the prominent peaks to be the TRGB of each galaxy.
The relatively weaker peaks in the brighter part of the TRGB are thought to be the tips from intermediate AGB populations or noise.
We derive our best estimates of the TRGB magnitudes and their measurement errors from ten thousand simulations of bootstrap resampling. Derived values are $I=27.67\pm0.05$ for NGC 4038/39 and $I=27.77\pm0.04$ for NGC 5584.\\

\begin{deluxetable*}{lcc}
\setlength{\tabcolsep}{0.05in}
\tablecaption{A summary of TRGB distance estimates of NGC 4038/39 and NGC 5584}
\tablewidth{0pt}

\tablehead{ \colhead{ } & \colhead{NGC 4038/39} &  \colhead{NGC 5584}
}
\startdata
TRGB magnitude, $I_{TRGB}$			& $27.67\pm0.05$ 	& $27.77\pm0.04$   \\
Median TRGB Color, $(V-I)_{TRGB}$ 	&$1.62\pm0.02$  	& $1.66\pm0.02$    \\
Galactic extinction, $A_I$, $E(V-I)$ & 0.06, 0.05& 0.06, 0.05  \\
Intrinsic TRGB magnitude,  $I_{0,TRGB}$ & $27.61\pm0.05$ & $27.71\pm0.04$ \\
Intrinsic TRGB color, $(V-I)_{0,TRGB}$ & $1.57\pm0.02$ & $1.61\pm0.02$ \\
Absolute TRGB magnitude, $M_{I,TRGB}$ 	& $-4.06$ 	& $-4.05$   \\
Distance Modulus, $(m-M)_0$ 	& $31.67\pm0.05\pm0.12^a$ 	& $31.76\pm0.04\pm0.12^a$   \\
Distance, d 			& $21.58\pm0.50\pm1.19^a$ Mpc	& $22.49\pm0.41\pm1.24^a$ Mpc   \\
\tablenotetext{a}{Systematic errors.}
\enddata
\label{distance}
\end{deluxetable*}

 We adopted the TRGB calibration suggested by \citet{riz07}, $M_{I,TRGB} = -4.05+0.217[(V-I)-1.6]$, which was also used in our previous studies \citep{lee12, lee13, jan14}. Median TRGB colors estimated from 
 $\sim200$ RGB stars in the $\pm0.03$ magnitude range of the TRGB are $1.62\pm0.02$ for NGC 4038/39 and $1.66\pm0.02$ for NGC 5584.
 We adopted 
 foreground extinction values, $E(B-V)=0.037$ for NGC 4038/39 and $E(B-V)=0.035$ for NGC 5584 from \citet{sch11}. Host galaxy extinctions for the blue RGB stars are expected to be very small, so they are ignored.
 By combining the TRGB magnitudes, extinction values, and the TRGB calibration, we obtained values of the distance moduli, $(m-M)_0=31.67\pm0.05$ (random) $\pm0.12$ (systematic) for NGC 4038/39, and $(m-M)_0=31.76\pm0.04$ (random) $\pm0.12$ (systematic) for NGC 5584,
  as summarized in Table \ref{distance}.

We checked the effect of TRGB color selection to the TRGB distance estimates. We divided the RGBs into four (for NGC 4038/39) and three (for NGC 5584) subregions with different $V-I$ ranges. The TRGB colors of these subregions cover $1.2 < V-I < 2.4$ for NGC 4038/39 and $1.2 < V-I < 2.1$ for NGC 5584, in steps of 0.3 mag. We then derived the TRGB distances for each subregion by applying the same method as done for the blue RGB stars. The TRGB distances for four subregions of NGC 4038/39 show similar values: $(m-M)_0=31.60$ to 31.77 with uncertainties of 0.04 to 0.09. A weighted mean of these four estimates is $(m-M)_0=31.64\pm0.03$, which is in good agreement with the estimate from the blue RGB stars, $(m-M)_0=31.67\pm0.05$. The three subregions of NGC 5584 also show similar values: $(m-M)_0=31.71$ to 31.78 with uncertainties of 0.04 to 0.06. A weighted mean of these estimates is $(m-M)_0= 31.77\pm0.03$, showing an excellent agreement with the estimate from the blue RGB stars, $(m-M)_0= 31.76\pm0.04$. Therefore, we conclude that the TRGB distance estimation depends little on the color selection of RGB stars.

\section{Discussion}

\subsection{Comparison with Previous Distance Estimates}

Tables \ref{distance_4038} and \ref{distance_5584} summarize the distance estimates for each galaxy in this study and in previous studies. Distance estimates for NGC 4038/39 show a large range : $(m-M)_0=30.62$ to $31.74$.
\citet{rie11} presented a distance estimate of $(m-M)_0=31.66\pm0.08$ from the photometry of 35 Cepheid variables.
They derived periods of the Cepheid variables from the image data taken with the $HST$/WFPC2 
and apparent magnitudes from the $F160W$ image data taken with the $HST$/WFC3. 
Later, \citet{fio13} applied their theoretical model to the observational data in \citet{rie11}, and derived a similar value, $(m-M)_0=31.55\pm0.06$.

\citet{sav04, sav08} reported relatively short distance estimates based on the TRGB method: $(m-M)_0=30.70\pm0.25$ from the WFPC2 \citep{sav04}, and $(m-M)_0=30.62\pm0.17$ from the ACS image data \citep{sav08}.
However, \citet{sch08} reported a much larger value of the TRGB distance, $(m-M)_0=31.51\pm0.17$ from the same ACS image data as used in \citet{sav08}. 
This $\sim0.9$ mag difference comes from the different TRGB magnitudes: $F814W_{0,TRGB}=26.59\pm0.09$ in \citet{sav08} and $T_{RGB}=27.46\pm0.12$ in \citet{sch08}. 
\citet{sch08} derived even larger values for the distance moduli: $(m-M)_0=31.7\pm0.3$  from the peak luminosity of SN Ia, SN 2007sr, 
and $(m-M)_0=31.76\pm0.27$ from the systematic recession velocity relative to the Local Group, applying the large scale flow model by \citet{ton00}.
By considering the distance moduli from three independent methods, \citet{sch08} suggested to use a conservative value, $D=22 \pm3$ Mpc ($(m-M)_0=31.7\pm0.3$) as the distance to NGC 4038/39.

\begin{deluxetable*}{llccl}
\tabletypesize{\footnotesize}
\setlength{\tabcolsep}{0.05in}
\tablecaption{A summary of distance measurements for NGC 4038/39}
\tablewidth{0pt}

\tablehead{ \colhead{ID } & \colhead{References} &  \colhead{Method} &  \colhead{Distance Modulus} &  \colhead{Remarks}
}
\startdata
1	& \citet{ama10} & SN Ia & $30.92\pm0.36$ 	& SALT2, $H_0=74.2$ km s$^{-1}$ Mpc$^{-1}$\\
2	& \citet{sch08} & SN Ia & $31.74\pm0.27$ 	& $H_0=72$ km s$^{-1}$ Mpc$^{-1}$ \\ 
3 	& \citet{sch08} & Recession Velocity 	& $31.76\pm0.27$ & $H_0=72$ km s$^{-1}$ Mpc$^{-1}$ 	  \\
4 	& \citet{rie11} &Cepheid& $31.66\pm0.08$  	&  $\#$ Cepheids = 35\\
5 	& \citet{fio13} &Cepheid& $31.55\pm0.06$  	&  $\#$ Cepheids = 32\\
6 	& \citet{sav04} &TRGB 	& $30.70\pm0.25$  	& $I_{TRGB}=26.5\pm0.2$\\
7 	& \citet{sav08} &TRGB 	& $30.62\pm0.17$  	& $F814W_{0,TRGB}=26.59\pm0.09$\\
8 	& \citet{sch08} &TRGB 	& $31.51\pm0.17$  	& $T_{RGB}=27.46\pm0.12$\\
9 	& This Study 	&TRGB 	& $31.67\pm0.05$  	& $I_{TRGB}=27.67\pm0.05$\\

\enddata
\label{distance_4038}
\end{deluxetable*}

\begin{deluxetable*}{llccl}
\tabletypesize{\footnotesize}
\setlength{\tabcolsep}{0.05in}
\tablecaption{A summary of distance measurements for NGC 5584}
\tablewidth{0pt}

\tablehead{ \colhead{ID } & \colhead{References} &  \colhead{Method} &  \colhead{Distance Modulus} &  \colhead{Remarks}
}
\startdata
1	& \citet{the07}	& TF & $31.12\pm0.40$ &  Mean of $J, H, K$-band \\
2	& \citet{spr09}	& TF & $31.59\pm0.49$ &   \\
3	& 			 	& TF & $31.48\pm0.52$ &   \\
4	& \citet{cou12}	& TF & $31.36\pm0.40$ &  $I$-band \\
5	& \citet{lag13} & TF & $30.69\pm0.69$ &  $H_0=100.0$ km s$^{-1}$ Mpc$^{-1}$ \\
6	& 				& TF & $31.40\pm0.69$ &  $H_0=72.0$ km s$^{-1}$ Mpc$^{-1}$ \\
7	& \citet{sor12}	& TF & $31.41\pm0.42$ & $3.6\mu$m, $(m-M)_{0,LMC}=18.48$ \\
8	& \citet{sor14} & TF & $31.28\pm0.43$ &  $3.6\mu$m, $(m-M)_{0,LMC}=18.48$ \\
9	& \citet{bro10}	& SN Ia & $32.30\pm0.08$ & MLCS2k2, $R_V=1.7$, $H_0=72.0$ km s$^{-1}$ Mpc$^{-1}$ \\
10 	&  				&  SN Ia & $32.29\pm0.10$ & MLCS2k2, $R_V=3.1$, $H_0=72.0$ km s$^{-1}$ Mpc$^{-1}$\\
11	& \citet{bur11}	&	SN Ia	& $31.90\pm0.00$ & SNooPy, $H_0=72.0$ km s$^{-1}$ Mpc$^{-1}$\\
12	& \citet{ama10} & SN Ia & $31.34\pm0.37$ & SALT2, $H_0=74.2$ km s$^{-1}$ Mpc$^{-1}$\\
13 	& \citet{rie11}	&Cepheid & $31.72\pm0.07$   &  $\#$ Cepheids = 94\\
14 	& \citet{fio13} &Cepheid& $31.71\pm0.04$  	&  $\#$ Cepheids = 78\\
15 	& This Study 	&TRGB & $31.76\pm0.04$   & $I_{TRGB}=27.77\pm0.04$\\

\enddata
\label{distance_5584}
\end{deluxetable*}

Our best estimate of the TRGB distance to NGC 4038/39 is $(m-M)_0=31.67\pm0.05$.
It is roughly one mag larger than that found in \citet{sav08}. 
Our estimate is not much different, within errors, from the TRGB distance given by \citet{sch08} ($(m-M)_0=31.51\pm0.17$). 
We investigate any cause for the small difference of 0.16 mag between our best estimate and that of \citet{sch08}. There are three differences in the methods between the two studies. First, \citet{sch08} used relatively redder RGB stars ($1.6<(V-I)_0<2.6$) than those used in this study. Second they used the entire ACS/WFC field, including tidal tail regions, while we used only the region with low sky values, avoiding tidal tail regions. Third, \citet{sch08} applied T magnitude (=$I_0-0.20[(V-I)_0-1.5]$), while we used I magnitude.
Using the same T magnitude as used in \citet{sch08}, we derived the TRGB magnitude of $T_{RGB}=27.47\pm0.05$, which is almost the same as the value given by \citet{sch08} ($T_{RGB}=27.46\pm0.12$). When we used the stars in the lowest sky background regions, avoiding tidal tail regions, as shown in Figure \ref{fig_finder} (a), we obtained a 0.05 mag fainter value, $T_{RGB}=27.52\pm0.05$. When we used the blue RGB stars ($1.2 \lesssim V-I\lesssim2.0$, inside the slanted box of Figure \ref{fig_cmd}) in the lowest sky background regions, we derive another 0.05 mag fainter value, $T_{RGB}=27.57\pm0.05$. The corresponding distance modulus is $(m-M)_0=31.62\pm0.05$, which is in agreement with our best estimate, $(m-M)_0=31.67\pm0.05$,  within an uncertainty. Thus, within the 0.16 magnitude difference between our best estimate and that of \citet{sch08}, 0.06 mag comes from field selection, 0.05 mag comes from the TRGB color selection, and 0.05 mag comes from the measurement error.
Our estimate is consistent with the Cepheid estimate by \citet{rie11} ($(m-M)_0=31.66\pm0.08$).

Distance estimates for NGC 5584 in the previous studies also show a large range from $(m-M)_0=31.12$ to $32.30$.
Distance estimates using the Tully Fisher relation give a large range ($(m-M)_0=31.12\sim31.59$) with large uncertainties \citep{the07,spr09,cou12,sor12,lag13,sor14}.
SN Ia distance estimates 
show a much larger range from $(m-M)_0=31.40$ to $32.30$, depending on the selection of the light curve fitting tool, the total to selective extinction ratio ($R_V$), and the value of the Hubble constant.
\citet{rie11} found 94 cepheid variables from the $HST$/WFC3 image data and derived a distance modulus of $(m-M)_0=31.72\pm0.07$.
Distance modulus derived in this study, the first distance estimate using the TRGB method, is $(m-M)_0=31.76\pm0.04$. 
This is in 
good agreement with the Cepheid estimate by \citet{rie11}, considering measurement errors.
To date, NGC 5584 is the most distant galaxy among the galaxies to which distances were measured using the TRGB method.

\subsection{The Calibration of SNe Ia and the Hubble Constant}

\begin{deluxetable*}{llcccccc}
\tabletypesize{\footnotesize} 
\setlength{\tabcolsep}{0.05in}
\tablecaption{A summary of optical maximum luminosity calibration of SNe Ia}
\tablewidth{0pt}
\tablehead{ \colhead{Galaxy } & \colhead{SN Ia} & \colhead{Filters} & \colhead{$\Delta^a$} & \colhead{$A_V$$^a$} & \colhead{$m_V^0$$^a$} &  \colhead{Distance Modulus} &  \colhead{$M_V^b$} \\
 & & & & & & TRGB &
}
\startdata

NGC 4038/39		& SN 2007sr	& uBVri & $-0.086$ & $0.349$ & $ 12.404$ & $31.67\pm0.05$ & $-19.266\pm0.176$\\ 
NGC 5584		& SN 2007af	& uBVri & $-0.038$ & $0.346$ & $ 12.810$ & $31.76\pm0.04$ & $-18.950\pm0.171$\\ 
M101 (NGC 5457)	& SN 2011fe	& UBVRI & $-0.154$ & $0.157$ & $  9.955$ & $29.30\pm0.01$ & $-19.345\pm0.162$\\ 
M66 (NGC 3627)	& SN 1989B	& UBVRI & $ 0.060$ & $1.242$ & $ 10.591$ & $30.12\pm0.03$ & $-19.529\pm0.217$\\ 
M96 (NGC 3368)	& SN 1998bu	& UBVRI & $-0.021$ & $1.025$ & $ 10.793$ & $30.15\pm0.03$ & $-19.357\pm0.199$\\ 

\multicolumn{7}{l}{Weighted mean of five SNe}&$-19.266\pm0.081$\\
\multicolumn{7}{l}{Weighted mean of three low-reddened SNe (SN 2007sr, SN 2007af, and SN 2011fe)}&$-19.192\pm0.098$\\
\tablenotetext{a}{$\Delta$, $A_V$, and $m_V^0$ denote the luminosity/light-curve shape parameter, the host galaxy extinction, and the corrected apparent peak magnitude described in \citet{jha07}, respectively.}
\tablenotetext{b}{We added 0.08 mag as an MLCS2k2 fitting uncertainty, 10\% of the Milky Way and the host galaxy extinction value as an extinction uncertainty, and 0.14 mag as an intrinsic luminosity dispersion of SNe Ia.}

\enddata
\label{luminosity}
\end{deluxetable*}

The TRGB distance estimates of SNe Ia host galaxies in this study and our previous studies are very useful in determining the peak luminosity of SNe Ia and the Hubble constant.
M101 (NGC 5457) is a nearby face-on spiral galaxy hosting one SN Ia, SN 2011fe, the nearest low-reddened SNe Ia with modern photometry. It was discovered just 4 hours after its explosion, so complete monitoring observations have been made. Thus, it is an ideal target for the calibration of SNe Ia.
\citet{lee12} investigated nine $HST$/ACS fields around the inner region of M101 and derived a TRGB distance of $(m-M)_0=29.30\pm0.01$ (random)$\pm0.12$ (systematic).

M66 (NGC 3627) and M96 (NGC 3368) are nearby spiral galaxies in the Leo I group hosting SNe Ia, SN 1989B and SN 1998bu, respectively. 
The host galaxy extinction values of these two SNe Ia are 
known to be relatively high ($A_V\gtrsim1$). However, since they are nearby and 
their optical light curves from before their maximum magnitudes are available, they are still useful for the calibration of SNe Ia.
\citet{lee13} presented new TRGB distances to M66 and M96 based on the deep $HST$/ACS images : $(m-M)_0=30.12\pm0.03$(random)$\pm0.12$(systematic) for M66 and $(m-M)_0=30.15\pm0.03$ (random)$\pm0.12$ (systematic) for M96.
We included the TRGB distance estimates for these three galaxies in the luminosity calibration of SNe Ia as well as the two galaxies in this study, NGC 4038/39 and NGC 5584.

We compiled optical light curves of five SNe Ia in the following literature: 
uBVri photometric data for SN 2007sr and SN 2007af from the Carnegie Supernova Project\footnote{http://csp.obs.carnegiescience.edu/, \citep{ham06}} ($u$-band) and \citet{hic09} (BVri bands), 
UBVRI photometric data for SN 2011fe, SN 1989B and SN 1998bu from \citet{per13}, \citet{wel94}, and \citet{jha99}, respectively.
We added three $V$-band data points before the pre-maximum of SN 2007sr, given by the All-Sky Automated Survey (ASAS) and \citet{rie11}.
All the photometric data for these five SNe Ia were done with CCD detectors, not photoelectric detectors.
Then we ran MLCS2k2 (version 0.07) \citep{jha07} to derive their light curve parameters.
We set the total to selective extinction ratio, $R_V=3.1$ for the Milky Way and $R_V=2.5$ for SN Ia host galaxies, to be consistent with \citet{rie11}.
We added 0.08 mag as a light curve fit error, and 10\% of the Milky Way and host galaxy extinction values as extinction errors in the error propagation.
We also included the intrinsic luminosity dispersion of SNe Ia of 0.14 mag \citep{jha07} in the error calculation.

\begin{figure}
\centering
\includegraphics[scale=0.8]{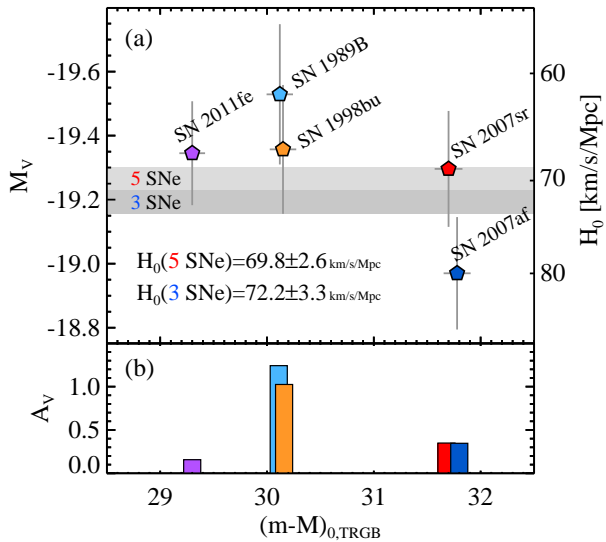} 
\caption{(a) Comparison of absolute magnitudes of five SNe Ia with their TRGB distances derived in this study and from our previous studies. Two shaded regions denote the weighted mean of five (upper region) and three (lower region) SNe Ia. (b) Comparison of $V$-band host galaxy extinction values of SNe Ia and their TRGB distances. Note that the extinction values for SN 1989B and SN 1998bu are much higher than other three SNe Ia.
}
\label{fig_5sne}
\end{figure}

The values of the parameters derived from the light curve fits are summarized in Table \ref{luminosity}.
Our light curve fit results are consistent with those in \citet{jha07} and \citet{rie11}, within fitting uncertainty. 
Figure \ref{fig_5sne} shows a comparison of TRGB distance estimates, $V$-band absolute peak magnitudes, and host galaxy extinction values of the five SNe Ia.
Two features are noted.
First, the host galaxy extinction values for SN 1989B and SN 1998bu are much larger than those of the other three SNe Ia. 
Thus the derived light curve parameters from these two SNe may be less reliable, although we corrected for their extinctions.
Second, SN 2007af is as bright as those in our previous studies, while SN 2007af is somewhat fainter.
The standard deviation for five SNe Ia is 0.21 mag, slightly larger than the luminosity dispersion of the distant SNe Ia, 0.14 mag \citep{jha07}.
This may be due to small number statistics.
Further studies adding more SNe Ia in the luminosity calibration is needed.

A weighted mean value of the $V$-band absolute peak magnitudes of five SNe is estimated to be $M_V=-19.27\pm0.08$.
It is $0.15$ mag brighter than that of the SH0ES project ($M_V=-19.12$, \citet{rie11}) and 0.2 mag fainter than those of the Hubble Key project ($M_V=-19.46$, \citet{fre01}) and the $HST$ key project ($M_V=-19.46$, \citet{san06}). 
If we adopt 3 low-reddened SNe (SN 2007sr, SN 2007af, and SN 2011fe), then the absolute peak magnitude would be slightly fainter, $M_V=-19.19\pm0.10$, getting closer to the estimate given by \citet{rie11}.

\citet{rie11} presented the relation between the absolute magnitude of SNe Ia and the Hubble constant  based on the MLCS2k2 fits of 140 SNe Ia at $0.023 < z < 0.1$ :
\begin{displaymath}
log H_0 = \frac{M_V^0 + 5a_v + 25}{5} ~{\rm and}~   a_v = 0.697\pm0.002.
\end{displaymath}

Using the absolute magnitudes derived in this study to the above equation, we derived a Hubble constant : $H_0 = 69.8\pm 2.6 $ (random)$ \pm 3.9$ (systematic) km s$^{-1}$ Mpc$^{-1}$ from the five SNe, and $H_0 = 72.2\pm 3.3 $ (random)$ \pm 4.0$ (systematic) km s$^{-1}$ Mpc$^{-1}$ from the three low-reddened SNe.
Our estimate based on the five SNe is similar to the values from the CMBR analysis, $H_0 \simeq 69$ km s$^{-1}$ Mpc$^{-1}$ \citep{ben13, pla15} and not much different, within the errors, from those of  recent Cepheid calibrations of SNe Ia, $H_0 \simeq 74$ km s$^{-1}$ Mpc$^{-1}$ \citep{rie11, fre12}.
Our estimate based on the three low-reddened SNe is between the values from the two independent methods.

\section{Summary}

We present the TRGB distance estimates for SN Ia host galaxies NGC 4038/39 and NGC 5584 from the PSF photometry of archival $HST$ images. A summary of the main results in this study is as follows.

\begin{itemize}

\item We detected RGB stars in the outer region of NGC 4038/39 and NGC 5584 based on the combined archival ACS/WFC and WFC3/UVIS image data. 

\item We derived photometric transformation between $F555W$ and $F814W$ bands of  WFC3/UVIS and that of ACS/WFC, by comparing the photometry of the resolved stars in two Milky Way globular clusters, NGC 2419 and 47 Tuc.

\item We found $I$-band TRGB magnitudes to be at $I_{TRGB} = 27.67\pm0.05$  for NGC 4038/39 and 
 $I_{TRGB}= 27.77\pm0.04$ for NGC 5584. 
 From these, we derived the TRGB distance estimates of $(m-M)_0=31.67\pm0.05\pm0.12$ for NGC 4038/39 and $(m-M)_0=31.76\pm0.04\pm0.12$ for NGC 5584.

\item Based on distance estimates for the two SNe derived in this study and that of the three SNe in our previous studies, we calibrate the $V$-band peak absolute magnitude of SNe Ia.

\item From the mean absolute magnitude of five SNe Ia, the  Hubble constant is determined to be $H_0 = 69.8\pm2.6\pm3.9$ km s$^{-1}$ Mpc$^{-1}$, which is similar to the values from the CMBR analysis, $H_0 \simeq 69$ km s$^{-1}$ Mpc$^{-1}$ \citep{ben13, pla15}. Our estimate is slightly smaller, but agrees within errors, with the values from the recent calibrations of SNe Ia with the Cepheid variables, $H_0 \simeq 74 $ km s$^{-1}$ Mpc$^{-1}$ \citep{rie11, fre12}. 
If we adopt 3 low-reddened SNe, corresponding the Hubble constant is slightly increased, $H_0 = 72.2\pm 3.3 $ (random)$ \pm 4.0$ (systematic) km s$^{-1}$ Mpc$^{-1}$.

\end{itemize}

\bigskip

This work was supported by the National Research Foundation of Korea (NRF) grant
funded by the Korea Government (MSIP) (No. 2012R1A4A1028713).
This paper is based on image data obtained from the Multimission Archive at the Space Telescope Science Institute (MAST). Thanks to Brian S. Cho for his comments and suggestions for the manuscript.


\clearpage


\end{document}